\documentstyle[12pt,aasms4]{article}

\begin{document}

\title {CO, \ion{C}{1} and \ion{C}{2} observations of NGC 7023}

\author {Maryvonne Gerin,\altaffilmark{1,2} Thomas G.
  Phillips,\altaffilmark{2} Jocelyn Keene,\altaffilmark{2} A.L.
  Betz,\altaffilmark{3} and R.T. Boreiko\altaffilmark{3} }

\altaffiltext{1}{Radioastronomie millim\'etrique, Laboratoire de
  Physique de l'ENS, 24 Rue Lhomond, 75231 Paris cedex 05, France}
\altaffiltext{2}{Caltech Submillimeter Observatory, Caltech 320-47,
  Pasadena, CA 91125}
\altaffiltext{3}{Center for Astrophysics \& Space Astronomy, Campus Box
  593, University of Colorado, Boulder, CO 80309-0593}

\abstract {We present new data on the photodissociation regions
  associated with the reflection nebula NGC~7023, particularly the
  three bright rims to the north, south and east of the illuminating
  star HD 200775.  $^{13}$CO(3--2) emission, mapped at 20\arcsec\ 
  resolution at the Caltech Submillimeter Observatory (CSO),
  delineates a molecular cloud containing a cavity largely devoid of
  molecular gas around this star. Neutral carbon is closely associated
  with the $^{13}$CO emission while ionized carbon is found inside and
  at the edges of the cavity. The ionized carbon appears to be, at
  least in part, associated with \ion{H}{1}. We have mapped the
  northern and southern rims in $^{12}$CO(6--5) emission and found a
  good association with the H$_2$ rovibrational emission, though the
  warm CO gas permeates a larger fraction of the molecular cloud than
  the vibrationally excited H$_2$.
  
  The column density contrast between the bright rims and the diffuse
  region inside and in front of the cavity is about 10. Despite the
  fact that the edges of the cavity are viewed edge-on, the carbon
  emission extends much further into the molecular gas than does the
  photodissociation region, as defined by the H$_2$ emission region.
  Geometrically, NGC 7023 consists of a sheet of dense molecular gas
  in which the star was born, subsequently blowing away much of the
  surrounding gas. The three bright rims are located at the edges of
  the remaining molecular cloud, and are viewed approximately
  edge-on.
  
  The results are compared with PDR models, invoking direct
  illumination from the star, which are largely successful, except in
  explaining the presence of neutral carbon deep in the molecular
  cloud. We suggest that, in the particular case of NGC~7023, a second
  PDR has been created at the surface of the molecular cloud by the
  scattered radiation from HD 200775. This second PDR produces a layer
  of atomic carbon at the surface of the sheet, which increases the
  predicted [C]/[CO] abundance ratio to 10\%, close to the observed
  value. Further tests for the applicability of PDR models in such
  regions are suggested.  }

\section{Introduction}

Reflection nebulae are produced when a massive star illuminates a
molecular cloud. Due to the enhanced radiation field, the gas
surrounding the star is heated and its chemistry is modified.
Reflection nebulae belong to the generic class of sources known as
PDRs (Photon-Dominated Regions).  These are typically located at the
transition layer between warm, ionized gas produced by an intense
radiation field and a cold neutral atomic or molecular cloud. If this
definition is extended to low values of the radiation fields, down to
the average intensity of the interstellar radiation field (ISRF), $I =
G_0 \times \rm ISRF$ with $G_0 = 1$ at the solar radius, PDRs comprise
a very significant fraction of the mass of the neutral gas in our
Galaxy (Hollenbach \& Tielens 1995).  Furthermore, due to their high
efficiency in producing intense emission (line and continuum), PDRs
produce a significant fraction of the dust and gas emission of
external galaxies including the CO rotational lines and \ion{C}{1} \&
\ion{C}{2} fine structure lines (Bennett et al.\ 1994, Hollenbach \&
Tielens 1995).

In PDRs at the surface of molecular clouds illuminated by a strong
radiation field, $G_0$ $\simeq$ 1000 or more, the gas temperature is
high and, in general, atomic and molecular lines are intense. For some
reflection nebulae, the geometry of the cloud is visually revealed.
Such sources are thus suitable places to study basic physical
processes of the interstellar medium.  In particular the processes
leading to the thermal balance of interstellar gas can be investigated
in detail due to the knowledge of the radiation field and the overall
geometry of the source.

Carbon plays a key role in PDR models because it is expected to be
dominantly in three different species throughout the PDR\@. There is a
layer of almost totally ionized carbon at the outer edge, an
intermediate region where carbon is neutral and then the cold
molecular interior where carbon should be locked into CO\@. Because
oxygen and nitrogen remain neutral in the outer edge, no equivalent
structure can be seen in the main repositories of these other abundant
species so carbon plays a unique role in testing PDR models.  The
possibility of observations of C$^+$, C and CO at the same spatial
locations in the same source therefore provides a unique diagnostic of
predictions: the depth and width of the CI emission zone can test the
accuracy of the chemistry and radiative transfer, and the excitation
of C and CO are sensitive to the thermal balance.  Previous
observations of interstellar clouds reveal that atomic carbon is
generally both overabundant and more widely distributed than simple
models predict (Phillips \& Huggins 1981, Keene et al.\ 1985). This
has led to questions concerning the homogeneity of the cloud resulting
in proposals for porous clouds with greater than expected UV
penetration (Phillips \& Huggins 1981, Stutzki et al.\ 1988). In that
case a PDR structure is expected at the surface of each substructure
in the porous cloud as long as a significant UV field is present ($G_0
\geq 1$ to produce \ion{C}{1} (Spaans 1996)).

A widely used test for models of the surface of the PDR is the H$_2$
rovibrational emission in the near infrared.  H$_2$ reveals details of
the external zone of the PDR, but because the emission is very
sensitive to both the UV field intensity and the gas density, the
extent of the H$_2$ emission is much reduced compared to that of the
three carbon species, even in the case of a porous cloud.

The combination of data from both the H$_2$ rovibrational emission and
the carbon budget therefore represents a powerful tool for
investigating PDRs, particularly at high spectral resolution when the
velocity structure can be resolved.  We present in this paper a new
$^{13}$CO (3--2) map of the reflection nebula NGC~7023 together with
selected data of C$^{18}$O(3--2), $^{12}$CO(6--5), neutral (\ion{C}{1})
and ionized (\ion{C}{1}) carbon emission. A $^{12}$CO(3--2) map has
been taken but is not shown, because the structure of the cavity
region is obscured by the lower column density foreground material.
Though the NGC~7023 reflection nebula has been extensively
investigated in the past, this is the first observation of neutral
carbon and the first study of \ion{C}{2} at high spectral resolution.

NGC 7023 is illuminated by the young B3Ve star HD 200775 ($\alpha(1950)
= 21^{\rm h}00^{\rm m}59\fs 7, \delta(1950) = 67\arcdeg 57\arcmin
55\farcs 5$), at a distance of 600 pc (Rogers, Heyer, \& Dewdney
1995).  The radiation field is enhanced by a factor of a few thousand
over the average value in the solar vicinity (Chokshi et al.\ 1988).
Due to this enhanced radiation field, the very small dust grains
experience large excursions in temperature and produce continuum
emission in the near IR, as first observed by Sellgren (1984) for this
and other reflection nebulae. The scattering properties of dust grains
have also been investigated in NGC 7023 (Murthy et al.\ 1993, Witt et
al.\ 1993).  Maps of the dust continuum emission in the far infrared
(Whitcomb et al.\ 1981, Casey 1991) have been taken at a spatial
resolution of about 1\arcmin. The dust temperature varies from 50 K in
the vicinity of HD 200775 to 20K in the molecular cloud. The decrease
in dust temperature is accompanied by an increase of the dust opacity
at 250 $\mu$m, indicating a rise of column density in the molecular
cloud surrounding HD 200775.

The structure of the molecular gas has been determined at low spatial
resolution (1\arcmin\ -- 2\arcmin) by Watt et al.\ (1984), Fuente et
al.\  (1992) and Rogers et al.\ (1995) by mapping in $^{12}$CO,
$^{13}$CO and C$^{18}$O.  The $^{13}$CO emission is found at the
border of the bright nebula, and delineates a cavity with an hourglass
shape roughly centered on HD 200775, in which the $^{13}$CO and
C$^{18}$O emission is extremely weak or not detectable.  This is seen
in Figure 1, which is our new high spatial resolution ($\sim$
20\arcsec) $^{13}$CO(3--2) map of most of the region. At this
resolution the bright inner edge of the $^{13}$CO emission traces the
nebulosity seen on the POSS plate.  Rogers et al.\ (1995) were also
able to isolate the \ion{H}{1} emission from the nebula from the
general, widespread \ion{H}{1} emission. They found \ion{H}{1}
emission from the cavity, mostly north of the star, and conclude that
there is global pressure equilibrium between the warm gas around the
star and the gas in the molecular cloud.  \ion{H}{1} is slightly
redshifted compared to the molecular gas, by about 2 km$\;$s$^{-1}$.
Fuente et al.\  (1996) obtained an \ion{H}{1} map with 10\arcsec\ 
resolution which showed that the neutral gas does not completely fill
the interior of the cavity, but accumulates on the inner edges, at the
surface of the molecular gas, as expected for a stratified PDR\@. The
absence of significant \ion{H}{1} emission in the largest lobe of the
cavity, west of the star, and the presence of background stars
indicates that we are dealing with a sheet of material in which the
star HD 200775 has created a nearly complete hole (Rogers et al.\
1995).

Fuente et al.\ (1993) have investigated, at high spatial resolution
($\sim$ 12\arcsec), the northern part of the nebula, 1\arcmin\
north-west of the star. They found a sharp ridge in $^{13}$CO and
C$^{18}$O emission, which we here call the north rim. They have shown
that the gas chemistry is affected by the UV radiation and that
abundance ratios such as [CN]/[HCN] and [HCN]/[HNC] are enhanced in
the gas affected by the UV\@. As deduced from HCN measurements, the
density in the gas is fairly high, about 10$^5$ cm$^{-3}$ in the
ridge.  Somewhat lower densities ($0.3 - 2 \times 10^4$ cm$^{-3}$) are
obtained from the excitation of CO and its isotopes.

Ionized carbon and atomic oxygen have been detected at low spectral
resolution by Chokshi et al.\ (1988) in the nebula, and they presented
the first PDR model for NGC 7023.  The PDR model has been refined by
Lemaire et al.\ (1996), who obtained images of the rovibrational lines
of molecular hydrogen with 1\arcsec\ resolution. The H$_2$ emission is
concentrated in narrow and long filaments, located mostly north of the
star. These filaments have also been detected in HCO$^+$ (Fuente et
al.\ 1996). Lemaire et al.\ (1996) show that the H$_2$ emission is not
uniform in the filaments but shows variations down to the 1\arcsec\
scale.  Both the high brightness and spatial structure of the H$_2$
emission are evidence that the illuminated gas is fairly dense,
$n_{\rm H} \simeq 10^5~ {\rm cm}^{-3}$ (Lemaire et al.\ 1996, Martini,
Sellgren, \& Hora 1997). There is a second H$_2$ front south of the
star, fainter than the northern front by a factor 2--3.  Similar
filaments are found in R and V-R pictures (Watkin, Gledhill, \&
Scarott 1991) as sources of Extended Red Emission (ERE). H$_2$
emission and ERE are coincident in both fronts (50\arcsec\ NW and
70\arcsec\ S) but have different small scale structure (Lemaire at al.\
1996). It is clear from Figure 1 that a third powerful PDR exists on
the east rim of the cavity.  We have initiated some studies here, but
in total there is little data available on this. It should be
investigated further in the future.

Because high $J$ CO lines and the carbon fine structure lines are
predicted to be strong by these PDR models, we have observed the $J$ =
3--2 lines of $^{12}$CO, $^{13}$CO and C$^{18}$O, the $J$ = 6--5 line
of $^{12}$CO and the 1--0 line of \ion{C}{1} in NGC 7023 with the
Caltech Submillimeter Observatory (CSO).  We also present Kuiper
Airborne Observatory (KAO) spectra of the \ion{C}{2} emission at high
spectral resolution.  We describe the observations in the next
section, then present the results and compare the line maps with the
H$_2$ images in order to obtain a more precise understanding of the
geometry of the nebula. Finally we discuss the validity of the PDR
models for this source.

\section{Observations} 

The CO and \ion{C}{1} observations were performed at the CSO in July
\& October 1995, and July \& October 1996.  The telescope is equipped
with SIS receivers operated in double-sideband (DSB) mode. The
observations were performed by position switching, with the reference
position (devoid of emission) located 30\arcmin\ north of the star
HD 200775, which was used as the map center. For the \ion{C}{1}
observations, we also used the position ($-40$\arcsec, 0\arcsec) as a
reference position, since previous data had shown it to be devoid of
emission.  The $^{12}$CO and $^{13}$CO maps were made in the ``on the
fly'' (OTF) mode, where the telescope continuously sweeps across the
sky and data are binned in a series of positions. A reference position
is taken at the beginning and the end of each row, and is used to
subtract the sky emission. We selected a sweep rate of 1\arcsec /sec
and a map step of 10\arcsec, resulting in 10 sec on source integration
time per point.  The spectra were analyzed with a 1024 channels
acousto-optic spectrometer (AOS) with a total bandwidth of 50 MHz and
an effective spectral resolution of about 100 kHz. The main beam
efficiencies of the telescope were 0.65, 0.53 and 0.45 at 345, 492 and
691 GHz respectively.  The angular resolution at the CSO is 20\arcsec\
at CO(3--2), 15\arcsec\ at \ion{C}{1}(1--0) and 10\arcsec\ at CO(6--5).

The \ion{C}{2} 158 $\mu$m line was observed using the KAO during two
flights in August and November 1991, using a far-infrared heterodyne
receiver described in detail by Betz \& Boreiko (1993).  The local
oscillator was an optically pumped CH$_2$F$_2$ laser and the mixer was
a liquid nitrogen-cooled GaAs Schottky diode in a corner-reflector
mount. The back-end consisted of an AOS with 400 independent channels
each with 0.5 km$\;$s$^{-1}$ resolution. The system noise temperature
was 15,000 K SSB in August and 12,000 K SSB in November. The
diffraction-limited beam size on the KAO at 158 $\mu$m is 43\arcsec\ 
FWHM, and pointing accuracy is estimated to be 15\arcsec. Calibration
is performed relative to the Moon, for which a physical temperature of
395 K and emissivity of 0.98 are assumed. Overall, the \ion{C}{2}
antenna temperature should be accurate to within approximately 25\%,
comparable to the uncertainty in the CO and \ion{C}{1} observations.

\section {Overall geometry of NGC 7023}

\subsection {$^{13}$CO and $^{12}$CO data}

Figure 1 presents a map of the integrated intensity in the
$^{13}$CO(3--2) line; the contours represent the H$_2$ v = 1--0 S(1)
emission from Lemaire et al.\ (1996).  The $^{13}$CO map reveals a
cavity of hourglass shape around the star. Moreover, the emission is
not constant at the edges of this cavity but presents three maxima
located north-west, south and east of HD 200775, with peak antenna
temperatures of about 15 K. 

The three bright rims are located at the outer edge of \ion{H}{1}
maxima, and represent examples of PDRs with radiation fields relative
to the average radiation field proposed by Mathis, Mezger, \& Panagia
(1983, MMP) of $G_0 = 5000$ (north rim), 2000 (south rim) and 250
(east rim).  These numbers are computed for a B3 star with an
effective temperature of 17000 K (Chokshi et al.\ 1988, Casey 1991,
Draine and Bertoldi 1996), assuming that the distances are the
projected ones. Many slightly different values and color temperatures
for the average interstellar radiation field at 1000 \AA\ have been
proposed in the literature. These have been summarized by Draine and
Bertoldi (1996).  Among these, the MMP radiation field is most
appropriate because it has the same color temperature in the UV as
HD 200775 (17000 K).  The calculated fields are in reasonable agreement
with the fields deduced from H$_2$ near infrared emission seen from
the north and south rims.  There is no report of a search for H$_2$
emission from the east rim.

It appears that no large scale velocity gradient is present in this
cloud. There is however a gradual velocity shift along the north rim.

Figures 2 and 3 present representative spectra in $^{12}$CO(3--2),
$^{12}$CO(6--5), $^{13}$CO(3--2) and \ion{C}{1}(1--0). The selected
spectra were taken near the star position ($-10$\arcsec, 20\arcsec),
in the cavity (90\arcsec, 10\arcsec), in the bright rims: south
($-40$\arcsec, $-80$\arcsec), north ($-40$\arcsec, 30\arcsec)
($-40$\arcsec, 40\arcsec) and ($-50$\arcsec, 60\arcsec), and east
(210\arcsec, 10\arcsec) and in the molecular cloud ($-40$\arcsec,
90\arcsec).  The $^{13}$CO line profiles are nearly Gaussian with most
of the emission centered at the $V_{LSR} = 2.5 \rm ~km\;s^{-1}$
whereas the $^{12}$CO(3--2) profiles have complicated shapes with
multiple peaks or flat tops, mostly in the northern region of the map.
Towards the southern and northern rims, the $^{12}$CO(6--5) profiles
are nearly Gaussian with peak intensity of 30 -- 40 K. Whereas the
$^{12}$CO(6--5) line profile closely follows the the $^{12}$CO(3--2)
profile in the southern rim, it is considerably stronger than
$^{12}$CO(3--2) in the northern rim (compare position ($-40$\arcsec,
$-80$\arcsec) in the south rim with positions ($-40$\arcsec,
$-40$\arcsec) and ($-50$\arcsec, 60\arcsec) in the north rim).  This
difference between the $^{12}$CO(3--2) and (6--5) profiles is seen for
most of the observed positions located north of HD 200775, and even in
the molecular cloud away from the north rim. For example, towards the
position ($-40$\arcsec, 90\arcsec), the two lines do not peak at the
same velocity. A likely explanation for this behavior is the effect of
self absorption by cold foreground gas, which would affect mostly the
low J profiles.  The $^{13}$CO lines towards the eastern edge are
double peaked or somewhat broader than those towards the northern and
southern edges of the cavity. At the position (210\arcsec, 10\arcsec)
the two components visible in $^{13}$CO(3--2) line profile are merged
in the $^{12}$CO(3--2) profile. The $^{12}$CO(6--5) line is somewhat
narrower than the (3--2) line.

The main difference between the $^{13}$CO and the $^{12}$CO maps is
that $^{13}$CO shows the structure of the bright rims more clearly,
but is barely detected in the cavity, whereas the main isotope reveals
the presence of some diffuse molecular gas in the cavity and of
foreground material.

\subsection {The cavity}

In Figure 1, the $^{13}$CO emission is found at the outer border of
the H$_2$ emission, which itself is found further away from the star
than the K-band continuum radiation. This continuum radiation arises
both from stellar reflected light and from stochastically heated small
dust grains.  The overall absence of $^{13}$CO, and even \ion{H}{1}
and \ion{C}{1} (see below) around the star supports the hypothesis
that it has created a cavity of warm ionized gas around it, possibly
by the mechanical action of a past outflow as suggested by the opening
angle and straight edges. The rather low reddening of HD 200775, 1.5
mag  (Finkenzeller \& Mundt 1984), the presence of background stars
in the western lobe of the cavity and the absence of strong $^{13}$CO
emission from the back side suggest that the cavity was created in a
sheet of molecular gas, which has been essentially destroyed close to
the star. In the following we deduce more information about this
cavity and its edges from the analysis of the $^{13}$CO and $^{12}$CO
data and their comparison with the \ion{H}{1} and near infrared H$_2$
maps.

\subsection {The region around HD 200775}

The star HD 200775 has a color excess E(B-V) = 0.45, and a reddening of
1.5 mag (Finkenzeller and Mundt 1984, Buss et al.\ 1994).  It also
presents a CH absorption line in its visible spectrum, with a CH
column density of $3.2 \times 10^{13}$ cm$^{-2}$ (Federman et al.\
1994; 1997). The column densities of atomic and molecular hydrogen
have been measured by Buss et al.\ (1994) from UV absorption lines to
be $N({\rm H}) \sim 1.3 \times 10^{21}~{\rm cm}^{-2}$ and $N{\rm (H_2)
  \sim 3.5 \times 10^{20}~ cm^{-2}}$.  The value of the color excess
would indicate a somewhat higher value of the total column hydrogen
column density, namely $N({\rm H}) + 2N(\rm H_2) = 2.6 \times 10^{21}
~cm^{-2}$.  However, this atomic hydrogen is probably not totally
associated with the nebula, but is partly distributed along the line
of sight since the interferometrically measured \ion{H}{1} column
density at the position of HD 200775 in velocities associated with the
nebula is only about $5 \times 10^{20}$ cm$^{-2}$ (Fuente et al.\
1996).  We adopt a value for the total hydrogen column density to the
star of $2.0 \times 10^{21}$ cm$^{-2}$.  Federman et al.\ (1997) have
measured a velocity difference between the atomic absorption lines and
the molecular absorption lines: the former appear blueshifted relative
to the latter, which are seen at $V_{LSR} = 1 - 2 ~{\rm km\; s}^{-1}$.
The sole exception is the absorption due to the molecular ion CH$^+$
which has the same velocity as the atomic lines ($-2$ km$\;$s$^{-1}$).

We could not detect \ion{C}{1} and were barely able to detect
$^{13}$CO in the cavity, but $^{12}$CO(3--2) lines were seen
everywhere.  In the cavity as well as towards the north rim, the
$^{12}$CO(3--2) profiles are broad and complex with two or three
maxima. This complex shape is also seen in lower resolution (1--0) data
(Rogers et al.\ 1995) and probably results from the combination of
velocity structure and self absorption. Around the star, the
$^{12}$CO(3--2) profiles get smoother and the central velocity shifts
from $\sim 2.5$ km$\;$s$^{-1}$ to $\sim 1.5$ km$\;$s$^{-1}$.  The antenna
temperature is about 6 K, whereas the $^{13}$CO(3--2) peak intensity is
about 0.8 K.  This $^{12}$CO emission is probably associated with the
material which causes the CH and C$_2$ visible absorptions.  Indeed,
the $^{12}$CO(3--2) spectra toward the north rim present a wing at low
velocities (0--1 km$\;$s$^{-1}$), which is not seen in the $^{12}$CO(6--5)
line nor in the $^{13}$CO(3--2) or \ion{C}{1} lines. This material at
low velocity may be the continuation of the molecular gas seen in
absorption along the line of sight to HD 200775.

To find the physical conditions of the gas along the line of sight to
the star, we used an LVG model and adopted a kinetic temperature of 30
K, deduced from the peak intensity of $^{12}$CO lines, (3--2) and
(1--0), in the south part of the molecular cloud where the lines have
Gaussian shapes. This assumes that all the molecular gas lies at about
the same distance from the star and that the kinetic temperature of
the molecular gas is fairly homogeneous. The $^{12}$CO data around the
star (at positions (0\arcsec, 0\arcsec) and ($-10$\arcsec, 20\arcsec))
are well fitted with a gas density of $\simeq$ 3,000 cm$^{-3}$ and a
column density $N({\rm CO}) = 3.0 \times 10^{16}$ cm$^{-2}$. The
excitation temperature of the 3--2 transition is 13 K, and the opacity
is of the order 2.5.  This column density must be viewed as a lower
limit since the $^{12}$CO lines may still suffer from self absorption.
Using the same density and temperature, the weak $^{13}$CO emission
observed at these positions corresponds to a column density of about
$1.5 \times 10^{15}$ cm$^{-2}$, a value consistent with previous
determinations by Rogers et al.\ (1995).  The measured isotopic ratio
[$^{12}$CO]/[$^{13}$CO] is then larger than 20, well within the
expected range for diffuse illuminated gas where carbon fractionation
is thought to take place (see below, \S5).  Using the hydrogen column
density of $2.0 \times 10^{21}$ cm$^{-2}$ derived from the star
reddening and the CH absorption, the CO relative abundance to H is
$1.5 \times 10^{-5}$ in the gas along the line of sight to the star.
The $^{13}$CO abundance relative to H is $7.5 \times 10^{-7}$.  This
is smaller than the CO abundance relative to hydrogen in dark clouds,
$4 \times 10^{-5}$ (Irvine, Goldsmith, \& Hjalmarson 1987), but is
consistent with the presence of a thin sheet of molecular material
illuminated by the intense UV radiation of HD 200775 and located in
front of the nebula.

\subsection {Secondary maximum inside the cavity}

Inside the cavity at about 100\arcsec\ east of the star, a faint
maximum appears in the $^{13}$CO integrated intensity map (Fig. 1).
We present in Figures 2 and 3 the spectra obtained towards the
position (90\arcsec, 10\arcsec).  The line is centered at $V_{LSR} =
3.3 \rm ~km\;s^{-1}$, clearly different from the bulk velocity of the
molecular gas, 2.5 km$\;$s$^{-1}$.  Thus in this case the emission is
clearly real and not merely due to the error pattern of the telescope,
which is a common source of illusory features when looking at a low
emissivity region surrounded by high intensity emission.  The
$^{12}$CO(3--2) line profile is complex with multiple peaks. The
correspondence of the $^{13}$CO(3--2) peak at 3.3 km$\;$s$^{-1}$ with
a dip in the $^{12}$CO spectrum is a clear indication of self
absorption.  A second dip in the $^{12}$CO spectrum at 4.5
km$\;$s$^{-1}$ might be associated with a weak $^{13}$CO signal.  The
$^{12}$CO velocity component at 0 km$\;$s$^{-1}$, with $T_A^* = 2.5$ K
has no counterpart in $^{13}$CO.

Assuming again a kinetic temperature of 30 K, with a density of $1
\times 10^4$ cm$^{-3}$, the column density of the secondary maximum at
the position (90\arcsec, 10\arcsec) is $N(^{13}{\rm CO}) = 2 \times
10^{15}$ cm$^{-2}$. From the map published by Rogers et al.\ (1995),
the $^{13}$CO(1--0) peak temperature lies in the 3--4 K range,
consistent with the density and column density derived above. Due to
the self absorption in $^{12}$CO(3--2), which shows up in the
$^{12}$CO(1--0) spectra as well, it is not possible to derive a
$^{12}$CO column density for the 3.3 km$\;$s$^{-1}$ velocity component.
However, assuming a $^{13}$CO abundance relative to H$_2$ of $2 \times
10^{-6}$ at most, we can obtain a lower limit of the H$_2$ column
density of $N{\rm (H_2) \geq 1 \times 10^{21}~cm^{-2}}$.  An upper
limit is obtained by assuming the same abundance of $^{13}$CO relative
to hydrogen as towards the star where the gas is diffuse, $[^{13}{\rm
  CO}] = 7.5 \times 10^{-7}$, hence $N({\rm H + 2H_2}) \leq 3 \times
10^{21}~ {\rm cm}^{-2}$. A likely value is $N{(\rm H_2) = 1.5 \times
  10^{21}~cm^{-2}}$.  At this position the \ion{H}{1} column density
is about $4 \times 10^{20}~{\rm cm}^{-2}$, thus most of the gas along
this line of sight is molecular.  This gas could be either a remnant
of a dense clump, located inside the cavity and currently evaporating,
or be a remnant of the now disrupted cavity edge which used to extend
in front of or behind the cavity.

\subsection {The three bright rims}

We have observed $^{12}$CO(6--5) toward the bright rims. Figure 4
presents the contours of the $^{12}$CO(6--5) emission overlaid on the
H$_2$ picture from Lemaire et al.\ (1996) and spectra are displayed in
Figure 2. As observed for the $^{12}$CO(7-6) transition in other
reflection nebulae (Jaffe et al.\ 1990), the $^{12}$CO(6--5) lines are
extremely bright in NGC 7023 with peak intensities of 38 K in the
north rim. The profiles are Gaussian in the north, south, and east
rims (see the spectra at the positions ($-50$\arcsec, 60\arcsec)
($-40$\arcsec, $-80$\arcsec), \& (210\arcsec, 10\arcsec) displayed in
Fig 2.).

Overall, the $^{12}$CO(3--2) and (6--5) profiles are similar except
toward the north rim because of self absorption in the (3--2) line.
These intense CO lines must be produced in a warm medium, with kinetic
temperatures in excess of 30 K over most of the mapped region and
larger than 40 K at the $^{12}$CO(6--5) emission peaks. For both the
northern and southern rims, the $^{12}$CO(6--5) emission presents a
sharp edge on the side closer to the star, a maximum associated with
the H$_2$ rovibrational emission, and a shallow decrease going deeper
into the cloud. Because $^{12}$CO(6--5) emission is also detected
towards the east rim, it is interesting to ask whether H$_2$ emission
will also be observable at this rim, even though the value of $G_0$
should be much lower.
   
The $^{13}$CO(3--2) maximum temperatures vary from 6 to 17 K along the
edges of the cloud. With a kinetic temperature of 45 K, and densities
around 10$^4$ cm$^{-3}$ as inferred by Fuente et al.\ (1993) and
confirmed by us (see below), this corresponds to column densities in
the range $0.3 - 1.3 \times 10^{16}$ cm$^{-2}$.  Towards the south
rim, for example at position ($-40$\arcsec, $-80$\arcsec), we measured
N($^{13}$CO) = $6.5 \times 10^{15}$ cm$^{-2}$ for $T_K$ = 45 K and
$n{\rm (H_2) = 10^4~cm^{-3}}$, using the (1--0) data from Rogers et
al.\ (1995) to constrain the excitation.  In the northern and eastern
rims, the peak $^{13}$CO(3--2) temperature reaches 17 K. The column
density is then $N{\rm (^{13}CO) = 1.3 \times 10^{16}~cm^{-2}}$ at the
position ($-40$\arcsec, 50\arcsec) in the northern rim and $N{\rm
  (^{13}CO) = 1.1 \times 10^{16}~cm^{-2}}$ at the position
(210\arcsec, 10\arcsec) in the eastern rim. In these dense ridges, the
gas is totally molecular (apart from the carbon atom content) and the
$^{13}$CO fractional abundance should be the same as in dark clouds,
$1 - 2 \times 10^{-6}$ (Irvine et al.\ 1987, Wilson \& Rood 1994).
The total molecular gas column density is therefore $\simeq$ 10$^{22}$
cm$^{-2}$. Using the relationships between extinction and molecular
column densities established by Frerking, Langer, \& Wilson (1982),
Duvert, Cernicharo, \& Baudry (1986) and Lada et al.\ (1994), the
total extinction through the northern and eastern rims is 6 -- 10
magnitudes.  The higher value is preferred since, using the
C$^{18}$O(3--2) data, we obtain a column density of $N{\rm (C^{18}O) =
  2.7 \times 10^{15}~ cm^{-2}}$ at the position ($-40$\arcsec,
50\arcsec) in the northern rim, which indicates a visual extinction of
about 10 magnitudes using the same relationships as above. Comparing
to the gas in the cavity, the column density contrast is about a
factor 10. A summary of the physical parameters for these various
positions is given in Table 1.

In an edge-on view of a PDR, the H$_2$ rovibrational emission is
forced to lie at the edge of the C$^+$/C/CO region, although it does
not have to extend along the full length of the interface, because it
arises from regions of high density and requires a large UV intensity.
While the H$_2$ rovibrational emission closely follows the edge of the
molecular cloud on the north front, there is a clear discrepancy on
the south front. The CO(6--5) map (Fig. 4) follows more closely the
H$_2$ feature than either the CO(3--2) or the $^{13}$CO maps.  A likely
explanation is that the H$_2$ emission in the south comes from some
high density surviving clump in the cavity, at about the same distance
from the star as in the northern rim. The bulk of the molecular gas in
the southern rim itself is slightly farther from the star and so may
not achieve the necessary conditions for H$_2$ rovibrational emission.

\section {Carbon budget in the northern and southern fronts}

\subsection {C and C$^{18}$O} 

We have observed the ${^3P_1} - {^3P_0}$, 492 GHz fine structure line
of carbon along two cuts, at RA = $-40$\arcsec\ and Dec = +60\arcsec,
together with a few other points either observed by Fuente et al.\
(1993) or presenting strong H$_2$ v = 1--0 S(1) emission. Some
\ion{C}{1} spectra overlaid on $^{13}$CO(3--2) are shown in Figure 3.
The \ion{C}{1} emission clearly peaks at the interfaces between the
cavity and the molecular cloud with no line detected inside the
cavity, but with a strong extension into the molecular cloud.  By
contrast with $^{13}$CO and \ion{C}{1} which start to emit at the peak
of the H$_2$ emission, at RA = $-40$\arcsec, Dec = +30\arcsec\ in the
north rim, C$^{18}$O is found to start its emission deeper in the
cloud at RA = $-40$\arcsec, Dec = +50\arcsec.  C$^{18}$O was marginally
detected towards the southern rim. We present in Figure 5{\it a\/} a
comparison of the \ion{C}{1}(1--0), $^{13}$CO(3--2), C$^{18}$O(3--2)
and H$_2$ S(1) intensities along the two cuts, and show in Figure
5{\it b\/} the intensity ratios \ion{C}{1}/$^{13}$CO and
C$^{18}$O/$^{13}$CO\@.  The H$_2$ data have been smoothed to 15\arcsec,
the spatial resolution of the \ion{C}{1} data.

The \ion{C}{1}/$^{13}$CO(3--2) antenna temperature ratio is about 0.6
for the northern rim, with a sharp increase to 1.7 at the position of
the H$_2$ rovibrational emission. \ion{C}{1} is fainter compared to
$^{13}$CO(3--2) towards the south, with a \ion{C}{1}/$^{13}$CO(3--2)
antenna temperature ratio only reaching 0.4. Apart from this sharp
maximum at the cloud edge, \ion{C}{1} and $^{13}$CO behave in the same
way inside the cloud. As seen in previous studies of a variety of
clouds (e.g.\ Phillips \& Huggins 1981; Keene et al.\ 1985), the line
profiles are also very similar for both species (Fig. 3). The line
ratios have a somewhat larger scatter in the horizontal cut, with
\ion{C}{1}/$^{13}$CO(3--2) ranging between 0.5 and 1.3. The lower
value, 0.5, is found at offset -50\arcsec\ along the cut, a position
deep in the cloud with a large column density, because it shows both
the maximum C$^{18}$O(3--2) emission and
C$^{18}$O(3--2)/$^{13}$CO(3--2) intensity ratio. Indeed, at this
position, C$^{18}$O/$^{13}$CO reaches 0.2, the expected isotopic ratio
indicating that the gas is well shielded from the UV radiation and
consequent photodissociation. Along this horizontal cut, the
\ion{C}{1}/$^{13}$CO intensity ratio follows a pattern anticorrelated
with the C$^{18}$O/$^{13}$CO ratio, with largest value when isotopic
fractionation should be taking place. Thus the carbon line is seen
both at the edge of the PDR, as predicted by the models, but also
throughout the cloud as previously noted for S140 and M17 (Keene et
al.\ 1985). The only place where carbon is not detected is in the
cavity around the star.

We have calculated the carbon column densities using the LTE formula:
$ N({\rm C}) {\rm (cm^{-2})} = 1.9\times 10^{15} \int T_{MB}dv \times Q
e^{E_1/kT} $, where $Q$ is the partition function: $Q = 1 +
3e^{-E_1/kT} + 5e^{-E_2/kT}$, and $E_1 = 23.6$ K and $E_2 = 62.5$ K
are the energies of the fine structure levels of carbon.
Because both upper levels of the \ion{C}{1} ground triplet are easily
excited, LTE conditions are usually reached.  At a kinetic temperature
of 30 K, this gives:
\begin{displaymath}
N({\rm C}) {\rm (cm^{-2})} = 1.25 \times 10^{16} \int
T_{MB}dv \; ({\rm K\;km\;s^{-1}}).
\end{displaymath}
\noindent
Therefore the column density of carbon is $4.2 \times 10^{16}$
cm$^{-2}$ for the south rim ($-40$\arcsec, $-80$\arcsec), and $8.1
\times 10^{16}$ cm$^{-2}$ for the north rim ($-40$\arcsec, 30\arcsec).
At the same positions, the $^{13}$CO column densities are respectively
$6.5 \times 10^{15}$ cm$^{-2}$ and $1.6 \times 10^{15}$ cm$^{-2}$.
With a $^{12}$CO/$^{13}$CO abundance ratio variable between 20 and 70
to take into account possible fractionation effects, the possible
range of $^{12}$CO column density at these positions is $1.3 - 4.6
\times 10^{17}$ and $0.32 - 1.1 \times 10^{17}$ cm$^{-2}$.  Though
much of the carbon is locked into CO, the abundance of neutral carbon
relative to CO is still high, from at least 15\% up to 70\%.

As shown above, the \ion{C}{1}/$^{13}$CO(3--2) intensity ratio
decreases further inside the cloud and reaches a value of about 0.5.
We have looked at the positions ($-40$\arcsec, 90\arcsec) and
($-50$\arcsec, 60\arcsec) where $^{13}$CO, C$^{18}$O and \ion{C}{1}
have been detected. The observed intensity ratio corresponds to a
column density ratio $N($\ion{C}{1}$)/N(^{13}{\rm CO}) = 9$ and 6
respectively, and a C/CO abundance ratio of 12\% and 9\%, estimating
the CO column density from the $^{13}$CO and C$^{18}$O data with
isotopic ratios [$^{12}$CO]/[C$^{18}$O] = 500 and
[$^{12}$CO]/[$^{13}$CO] = 70. For those positions, the C$^{18}$O
column densities are $1.6 \times 10^{15}$ cm$^{-2}$ and $2.0 \times
10^{15}$ cm$^{-2}$ respectively with [$^{13}$CO]/[C$^{18}$O] column
density ratios of 6.0 and 6.5.  Thus even towards positions where most
of the gas does not experience heavy illumination, as evidenced by the
low value of the [$^{13}$CO]/[C$^{18}$O] column density ratio, neutral
carbon still represents about 10\% of the available gas phase carbon.
As discussed above, this is a common finding in illuminated clouds
(Keene 1995).
 
\subsection{C$^+$}

\ion{C}{2} observations have been taken with the KAO with a 45\arcsec\
beam using a heterodyne receiver (Betz \& Boreiko 1993).  We present
in Figure 6 a comparison of the \ion{C}{2} spectra with $^{12}$CO and
$^{13}$CO points averaged to represent the same beam on the sky.
Apparently, the \ion{C}{2} emission occurs with two separate
distributions. It has a component similar to $^{13}$CO with similar
centroid velocity for the positions covering the northern rim. For
positions centered in the cavity where we could barely detect
$^{13}$CO, \ion{C}{2} emission occurs at a more positive velocity.
This second component has clearly broader lines. Both the positive
velocity and the broader lines suggest that this emission may be
associated with the neutral hydrogen, which also appears at positive
velocities relative to the molecular gas (Rogers et al.\ 1995, Fuente
et al.\ 1996). The antenna temperature of the \ion{C}{2} line is about
10 K in the cavity and rises to 20 K for the position closest to the
H$_2$ edge. For a gas density larger than 10$^4$ cm$^{_3}$ and a
temperature of about 80 K, this corresponds to a column density of $1
- 3 \times 10^{17}$ cm$^{-2}$ (Jansen et al.\ 1996). However the
emission may not be uniform in the beam, since it is predicted from
PDR models to rise steeply at the H$_2$ front. It is clear that
\ion{C}{2} emission is associated with both the cavity and the PDR, in
contrast to \ion{C}{1} which is only seen in the PDR and the molecular
cloud.

The $^{12}$CO spectra in the cavity are centered at $V_{LSR} = 1 \rm
~km\;s^{-1}$ while the \ion{C}{2} line appears at a more positive
velocity, 3.5 km$\;$s$^{-1}$.  It is likely that the observed
\ion{C}{2} emission is associated with the \ion{H}{1} region inside
the cavity, while the $^{12}$CO and $^{13}$CO lines towards the cavity
are produced in the layer of molecular gas associated with the CH
absorption and located in front of the cavity.

\section{Comparison with PDR Models and Discussion}

PDR models are well established as capable of explaining the molecular
and atomic emission lines seen on the boundaries of molecular clouds
when illuminated by starlight (Hollenbach \& Tielens 1995). These
models have recently been modified to incorporate new data on reaction
rates and molecular excitation processes.  We have used the PDR model
based on the code of Abgrall et al.\ (1992) and Le Bourlot, Pineau des
For\^ets, \& Roueff (1993a), with different geometric configurations,
in order to check the validity of actual parameters and to investigate
the marked difference between the northern and the southern rims in
NGC 7023, in both the H$_2$ and \ion{C}{1} intensities relative to
$^{13}$CO\@. In part these differences could be due to geometrical
effects. Since this PDR model starts with a semi-infinite H$_2$ slab
on the surface of which the UV impinges, it cannot address the
structure of the cavity, the region between the PDR and the star.

>From Figure 5, it is clear that the north rim is viewed almost
edge-on: the large variation of the column density of molecular
hydrogen as traced by $^{13}$CO(3--2) emission, the position of the
molecular hydrogen emission at the edge of the molecular cloud, and
the coincidence of the peak in the \ion{C}{1}/$^{13}$CO ratio with the
zone of H$_2$ emission result naturally from the edge-on geometry. It
is quite likely that the actual geometry is that of a sheet with a
sharp cut-off edge, which is the surface illuminated by the star. The
sections of these edges closest to HD 200775 are the north and south
rims, where bright H$_2$ emission has been detected.  

In order to investigate the role of the viewing geometry in a PDR
model, we have calculated abundances as a function of depth within a
bright rim. These abundances are naturally interpreted as
corresponding to a ``face-on'' viewing geometry, and from them we
approximated the emergent intensities for a PDR viewed ``edge-on'' by
summing the local emissivities at each depth inside the cloud along a
path length of $3 \times 10^{16}$ cm (0.01 pc). This approximation is
valid only for optically thin lines for which the emissivities may be
enhanced significantly by limb brightening. The model we have used is
an isochoric model, with a UV enhancement factor $G_0$ of 10$^4$ and a
density of $n{\rm (H + 2H_2) = 2 \times 10^5~cm^{-3}}$.  We have used
as gas phase elemental abundances the values found towards $\zeta$ Oph
listed in Table 2. We have also tried a model with oxygen, carbon and
nitrogen depleted by a factor of three from these values.  From
Lemaire et al.\ (1996), it is known that the hydrogen density has to
be larger than about 10$^5$ cm$^{-3}$ to reproduce the observed
intensities of the H$_2$ rovibrational lines.  An analysis of the
molecular emission lines (CS and HCN) also suggest high densities, in
the 10$^5$ cm$^{-3}$ range (Fuente et al.\ 1993).

The results are illustrated in Figure 7 where we show the variation of
the local abundances $n({\rm X})/(n({\rm H}) + 2n({\rm H_2}))$ of
$^{12}$CO, $^{13}$CO, C and C$^+$ with the depth into the cloud
expressed in arcsec at the distance of NGC 7023 (600 pc) (top panel)
and the predicted emissivities for the edge-on view (bottom panel).
For the optically thick $^{12}$CO(3--2) line, little or no limb
brightening is expected and we show only the face-on emissivity. In
this plane-parallel structure, the model predicts a layered structure
with C$^+$ at the outer boundary of the cloud, then C and CO\@.  For
this isochoric model, in the outer region where C and C$^+$ dominate
over $^{12}$CO, the high density and temperature cause efficient
excitation of $^{12}$CO and therefore produces strong high $J$
$^{12}$CO emission.

We discuss below the different layers in the photo dissociation region:

1) \ion{C}{2} emission is found in the warm ionized and partially
dissociated medium which lies at the surface on which the UV impinges.
It is also present in the next layer where H$_2$ is able to resist
photodissociation and is strongly excited. This is qualitatively
consistent with the observed C$^+$ data which show two different
components depending on the actual observed positions. The emission
from the \ion{C}{2} region in the models would be clearly diluted in
the 45\arcsec\ KAO beam.  With a dilution factor of about 5, i.e.\
assuming that the \ion{C}{2} emission comes from a structure of width
10\arcsec\ and longer than 60\arcsec, the observed peak antenna
temperature of the \ion{C}{2} emission from this region corresponds to
an absolute antenna temperature of 75 K, close to the predicted
values.

2) The next layer is the atomic carbon region and the main problem for
the models is the fit of the carbon data: whereas the model predicts
no \ion{C}{1} emission at distances larger than 20\arcsec\ from the
PDR, \ion{C}{1} has been observed throughout the molecular cloud (Fig.
5). The distribution of neutral carbon is therefore much more extended
than expected for a simple PDR\@.  Because reflected starlight is seen
in this nebula, and neutral carbon is produced in even low radiation
field environments, it is possible that some neutral carbon emission
we observe comes from the externally illuminated visible surface of the
sheet.  We have evaluated the UV intensity due to scattered light from
HD 200775 using the observations with UIT (Witt et al.\ 1992).  We
found an enhancement factor of $G_0 \sim 20$ at 100\arcsec\ from HD
200775. This scattered radiation forms another PDR at the surface of
the molecular cloud. To evaluate the role of this second PDR in
producing neutral and ionized carbon at large distances from the north
rim, we have run a PDR model with the same abundances, the same high
density, $n = 2 \times 10^5$ cm$^{-3}$, but a low radiation field
$G_0$ = 50.  The variation of the abundances of C$^+$, C and CO as a
function of the depth into the cloud A$_V$ are displayed in Figure 8.
For A$_V$ of a few magnitudes, this second PDR is an important source
of neutral carbon in the molecular gas, so that the predicted [C]/[CO]
abundance ratio is larger than 10\%. Indeed, even in a low UV field,
the PDR is associated with ionized and neutral carbon layers at the
surface of the molecular cloud. When viewed ``face-on'' the layered
structure results in significant cumulative column densities of
neutral and ionized carbon for all values of A$_V$ hence significant
abundances of C$^+$ and C appear, even though most of the neutral and
ionized carbon reside at the outer surface. With this low UV flux, the
model does not predict any H$_2$ or warm dust emission, as observed.

In further studies this proposal could be tested by determining if a
substantial amount of C$^+$ is also present in the mixed C/CO
regime, which could indicate a UV origin to the effect. This would
require a platform such as SOFIA or FIRST with heterodyne receivers.
PDR models predict the existence of a zone where carbon isotopic
fractionation takes place at the edge of the molecular cloud close to
the UV source.  In that zone, the [$^{12}$CO]/[$^{13}$CO] abundance
ratio drops to 30 (Kopp et al.\ 1996).  Carbon fractionation takes
place as soon as the formation of molecules enhances the cooling and
the temperature drops.  It stops deeper into the cloud because of the
lack of C$^+$. If the PDR model is appropriate for these regions far
from the star, $^{13}$CO/$^{12}$CO fractionation should exist in such
a regime and could be observed with ground based techniques. If
carbon deeper into the cloud is due to non-UV processes, such as
turbulent mixing (Xie, Allen, \& Langer 1995) or high ionization phase
chemistry (Le Bourlot et al.\ 1993b; Le Bourlot, Pineau des For\^ets,
\& Roueff 1995; Flower et al.\ 1994), there will be much less C$^+$ and
fractionation.

The assumed values for elemental abundances are important parameters
for PDR models.  The models with depleted elements exhibit the same
brightness for the $^{12}$CO and $^{13}$CO lines essentially due to
the high optical depths, but reduced emissivities of ionized and
neutral carbon by a factor roughly equal to the input depletion.  The
strength of both the observed \ion{C}{1} and \ion{C}{2} lines favor
the models with rather high gas phase abundances of carbon and oxygen,
i.e.\ the models using the $\zeta$ Oph abundances.  Because the cooling
is slightly reduced in the models with lowered elemental abundances,
the $^{13}$CO edge is not as sharp as in the case of $\zeta$ Oph
abundances.  In any case, the edge is likely to be unresolved because
the emission rises over $3 \times 10^{16}$ cm. At the distance of NGC
7023, 450 -- 600 pc, this corresponds to at most 4\arcsec, which is
below the resolution of these observations but could be resolved by
millimeter interferometers.

\acknowledgements 

We thank J. Le Bourlot, G. Pineau des For\^ets and E. Roueff for the
use of their PDR model. The CSO is funded by NSF contract AST96-15025.
Work on the NASA Kuiper Airborne Observatory was carried out under NASA
grant NAG 2-605 to JBK.

\clearpage

\begin{deluxetable}{lrrrrrrr}
\tablecolumns{8}
\tablecaption{Physical parameters at representative positions}

\tablehead{ \colhead{Position} & \colhead{$T_K$} & \colhead{$n({\rm
      H_2})$} & \colhead{$N{(\rm ^{13}CO)}$} & \colhead{$N{(\rm
      C^{18}O)}$} & \colhead{$N$(C)} & \colhead{$\frac{[^{13}{\rm
          CO}]}{[{\rm C}^{18}{\rm O}]}$} & \colhead{$\frac{[{\rm C}]}
    {[{\rm CO}]}$}\\
  \colhead{arcsec} & \colhead{K} & \colhead{cm$^{-3}$} &
  \colhead{cm$^{-2}$} & \colhead{cm$^{-2}$} & & & }

\startdata
(0,0) & 30 & $2 \times 10^3$ & $1.1 \times 10^{15}$  & & & & \nl
(-10,20) & 30 & $10^4$ & $ 1.8 \times 10^{15}$ & & $\leq 4.0 \times 10^{16}$
 & & $\leq 0.3$ \nl
\tableline
(-40,-90) & 45 & $10^4$ & $5.4 \times 10^{15}$ & $13 \times 10^{14}$ & 
$4.7  \times 10^{16}$  & 4.0 & 0.12  \nl
(-40,-80) & 45 &  $10^4$ & $6.5 \times 10^{15}$ & $ 13 \times 10^{14}$ & 
$4.2  \times 10^{16}$ & 5.0 & 0.09 \nl
(-40,-70) & 45 & $7 \times 10^3$ & $3.7 \times 10^{15}$  & 
$ 6.0 \times 10^{14}$ & $4.0  \times 10^{16}$ & 6.2  & 0.15 \nl
\tableline
(-40,30) & 45 & $7 \times 10^3$ & $1.6 \times 10^{15}$ &
 $\leq 7.0 \times 10^{14}$ & $8.1  \times 10^{16}$  & $\geq 2$ & 0.70\nl
(-40,40) & 45 & $10^4$ & $7.7 \times 10^{15}$ & $14 \times 10^{14}$ 
& $14  \times 10^{16}$ & 5.5 & 0.26 \nl
(-40,50) & 55 & $10^4$ & $13 \times 10^{15}$ & $ 27 \times 10^{14}$ 
& $16  \times 10^{16}$ & 4.8 & 0.17  \nl
(-40,90) & 30 & $10^4$ & $9.6 \times 10^{15}$ & $ 16 \times 10^{14}$ 
&$7.8  \times 10^{16}$ & 6.0 & 0.12\nl
(-50,60) & 40 & $10^4$  & $13\times 10^{15}$    & $20 \times 10^{14}$
& $7.9  \times 10^{16}$ & 6.5 & 0.09 \nl
\tableline
(90,10) & 30 & $10^4$  & $2.2 \times 10^{15}$ & & & & \nl
(210,10) & 35 & $10^4$  & $ 11 \times 10^{15}$ & & & & \nl
\tablecomments{The column densities have been calculated with
  [$^{12}$CO]/[$^{13}$CO] = 70 and [$^{12}$CO]/[C$^{18}$O] = 500.}
\enddata
\end{deluxetable}

\clearpage

\begin{deluxetable}{lrl}
\tablecolumns{3}
\tablecaption{Elemental abundances used in the PDR model}

\tablehead{ \colhead{Parameter}  & \colhead{Value} & \colhead{Reference}}

\startdata
[O]/n$_{\rm H}$ & $2.93 \times 10^{-4}$ &  Savage et al.\ 1992 \nl
[C]/n$_{\rm H}$ & $1.32 \times 10^{-4}$ & Cardelli et al.\ 1993 \nl
[N]/n$_{\rm H}$ & $7.76 \times 10^{-5}$ &  Snow \& Witt 1996 \nl
\tableline
[$^{12}$C]/[$^{13}$C] & 70 \nl
[$^{16}$O]/[$^{18}$O] & 500 \nl
\enddata
\end{deluxetable}

\clearpage

\figcaption[]{Map of the $^{13}$CO(3--2) integrated area between 0 and
  4.5 km$\;$s$^{-1}$. The grey scale ranges from 1 to 12 $\rm
  K\;km\; s^{-1}$. Two contours of the H$_2$ v=1--0 S(1) (Lemaire et al.\
  1996) emission are overlaid (the circles around the star and the
  straight line features are artifacts of the IR data).}

\figcaption[]{Spectra observed in $^{12}$CO(3--2) (black line) and
  $^{12}$CO(6--5) (grey line) at representative positions in NGC 7023.
  Velocities are given relative to the LSR; positional offsets are in
  arc seconds relative to HD~200775, at $\alpha(1950) = 21^{\rm
    h}00^{\rm m}59\fs 7, ~\delta(1950) = 67\arcdeg 57\arcmin 55\farcs
  5 $.}

\figcaption[]{Spectra observed in $^{13}$CO(3--2) (black line) and
  CI (grey line) at representative positions in NGC 7023.
  Velocities are given relative to the LSR.}

\figcaption[]{Map of the $^{12}$CO(6--5) integrated intensity overlaid
  on the near infrared image of the H$_2$ v = 1--0 S(1) emission in
  NGC 7023 (Lemaire et al.\ 1996). The first contour is drawn at 14 
  $\rm  K\;km\; s^{-1}$, and the level spacing is 7 $\rm K\;km\; s^{-1}$. 
  Black dots show  the positions  observed in CO(6--5). }

\figcaption[]{ {\it a}\/) Comparison of the $^{13}$CO(3--2) (full line),
  CI (dashed line), C$^{18}$O(3--2) (bold line)
  and H$_2$ v = 1--0 S(1) (dotted line) emission along the two cuts at
  RA $= -40$\arcsec\ (top panel) and Dec $= +60$\arcsec\ (bottom
  panel).  The H$_2$ data have been smoothed to a resolution of
  15\arcsec\ to match the CI data.  {\it b}\/) Intensity ratios
  along the same cuts.}

\figcaption[]{Map of the CII spectra observed with the KAO with
  a 45\arcsec\ beam. The spectral resolution is 0.5 km$\;$s$^{-1}$. CO(3--2)
  spectra obtained by convolving the CSO data to the resolution of the
  KAO observations are shown with a dotted line.  $^{13}$CO(3--2)
  spectra obtained by convolving the CSO data to the resolution of the
  KAO observations are shown at high spectral resolution with a thin
  black line.  The unit for all spectra is $T_{A}^*$ in (K). Velocities
  are given relative to the LSR.}

\figcaption[]{{\it a}\/) Local abundances, $n({\rm X})/(n({\rm H}) +
  2n({\rm H_2}))$, as a function of the distance from the cloud edge
  for a PDR model with a total hydrogen density $n_{\rm H} = 2 \times
  10^5$ cm$^{-3}$, a UV field $G_0 = 10^4$ and elemental abundances
  from Table 2. We have used a distance of 600 pc to convert linear
  distances to angular scales.  {\it b\/}) Predicted emissivities
  (erg~cm$^{-2}$s$^{-1}$) of the PDR model for CI, CII,
  $^{12}$CO(6--5), $^{13}$CO(3--2), and $\rm H_2$ for an edge-on view.
  A depth of the slab along the line of sight of $3 \times 10^{16}$ cm
  (0.01 pc) is assumed.  The emissivity of the optically thick
  $^{12}$CO(3--2) line is shown for a face-on view.  The observed
  emissivities at the position ($-40$\arcsec, 50\arcsec) in the north
  rim are: $6.3 \times 10^{-7}$ erg~cm$^{-2}$s$^{-1}$ for
  $^{13}$CO(3--2) and $1.5 \times 10^{-6}$ erg~cm$^{-2}$s$^{-1}$ for
  CI\@.  For position ($-40$\arcsec, $-70$\arcsec) in the south rim,
  we found: $2.1 \times 10^{-7}$ erg cm$^{-2}$s$^{-1}$ for
  $^{13}$CO(3--2), $3.7 \times 10^{-7}$ erg cm$^{-2}$s$^{-1}$ for CI
  and $1.5 \times 10^{-6}$ erg cm$^{-2}$s$^{-1}$ for $^{12}$CO(3--2).}

\figcaption[]{Column density ratios $N({\rm X})/(N({\rm H}) + 2N({\rm
    H_2}))$ as a function of $A_V$ for a PDR model with the same
  density as Fig. 7, $n_{\rm H} = 2 \times 10^5$ cm$^{-3}$, and a
  lower UV field $G_0 = 50$.}

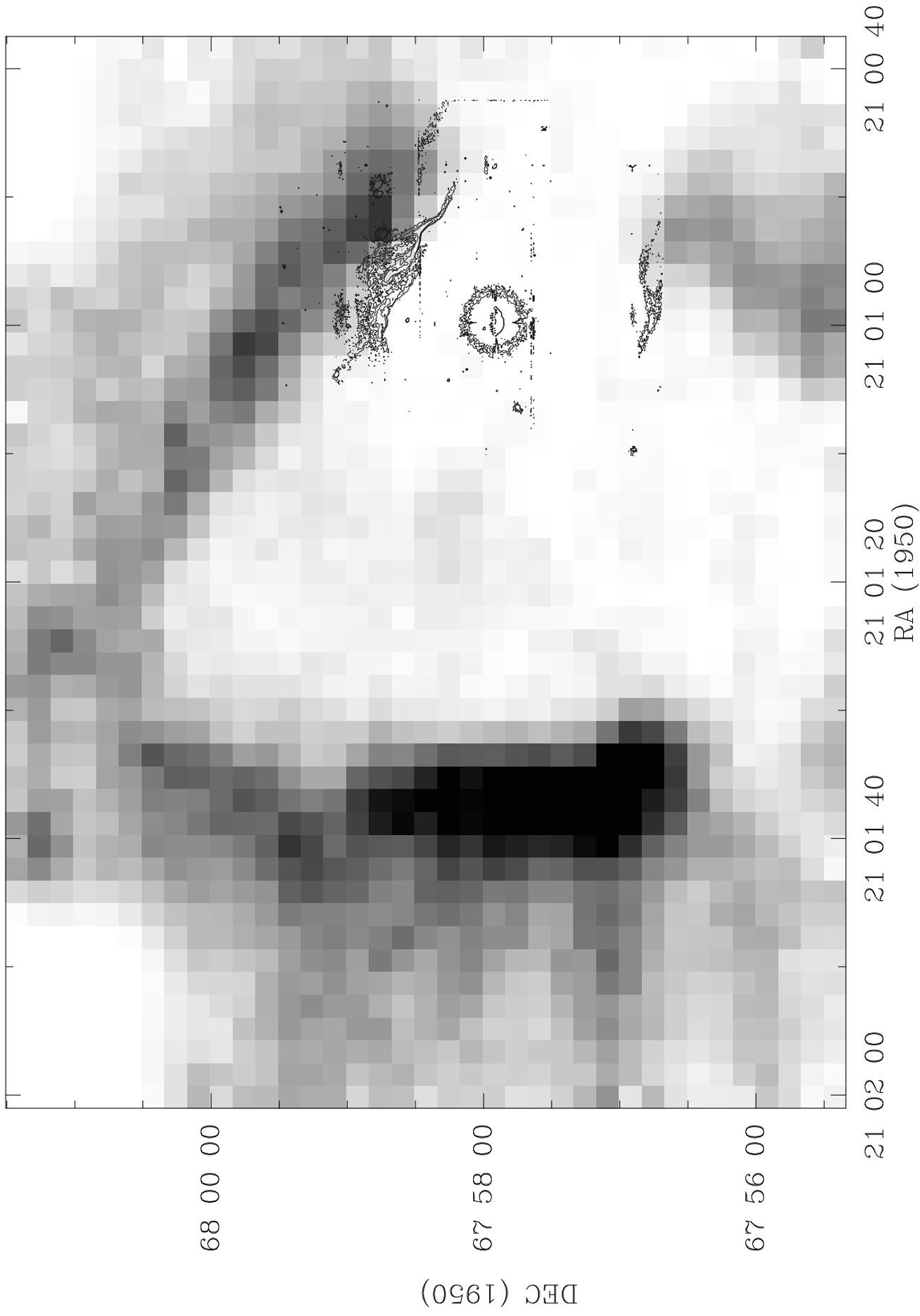
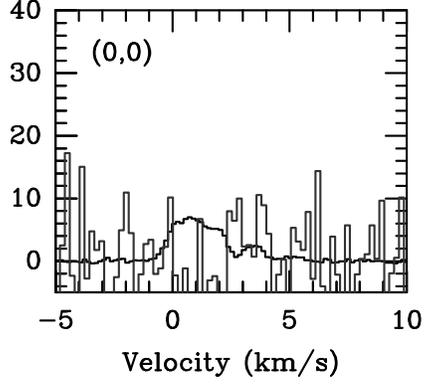
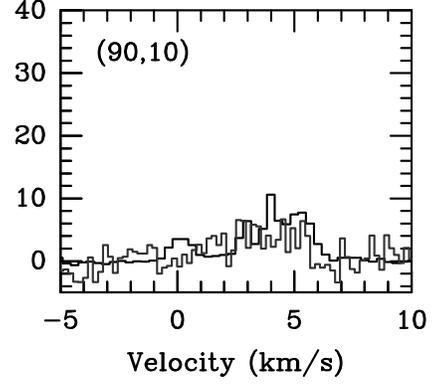
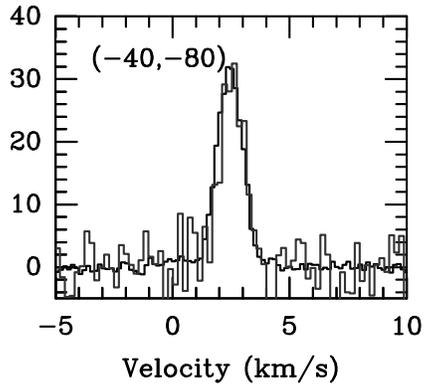
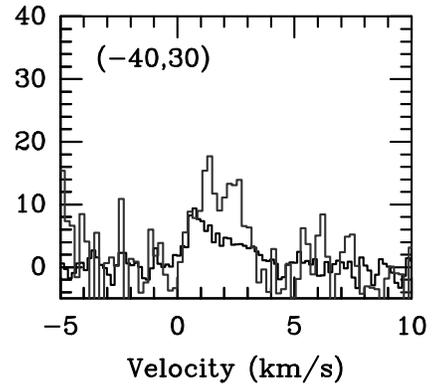
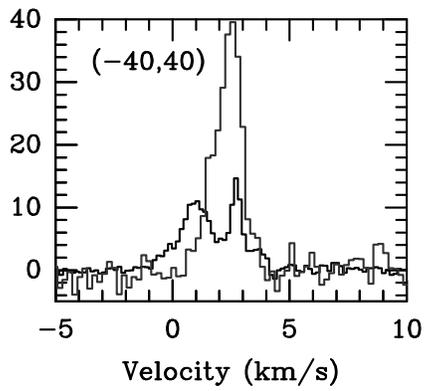
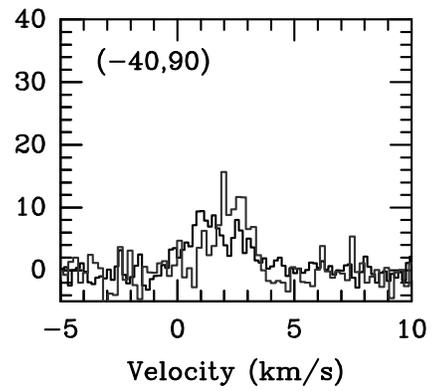
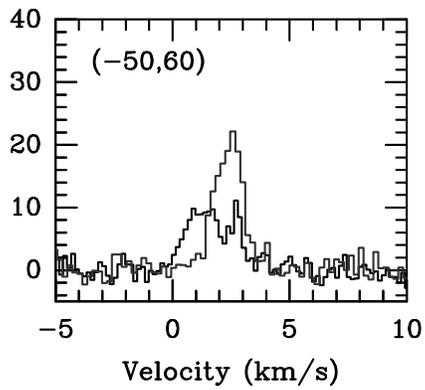
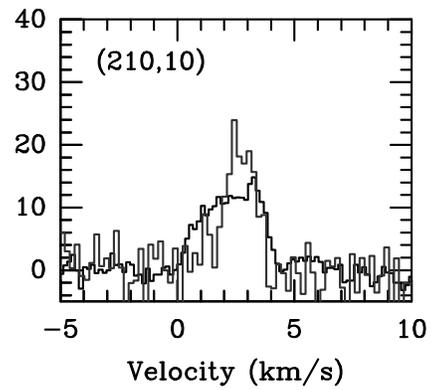
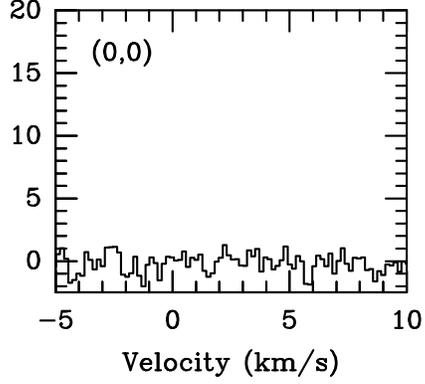
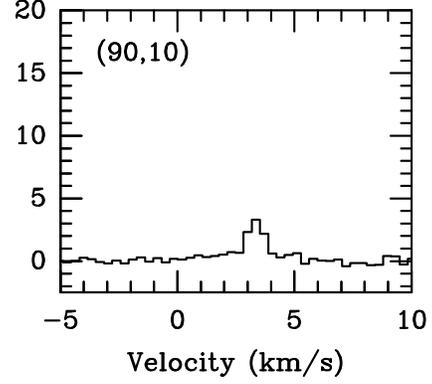
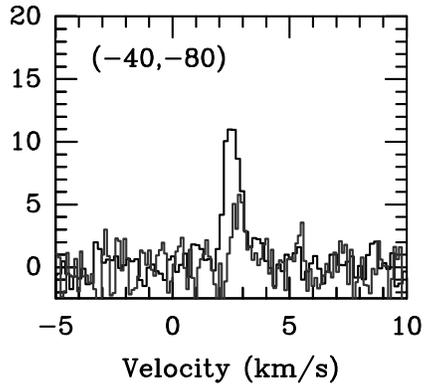
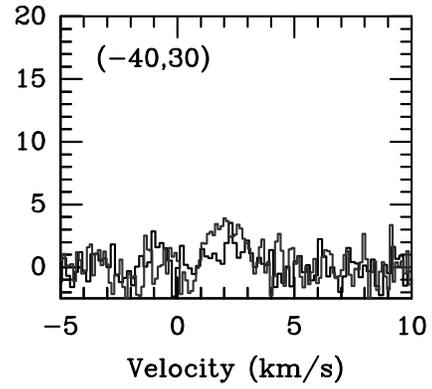
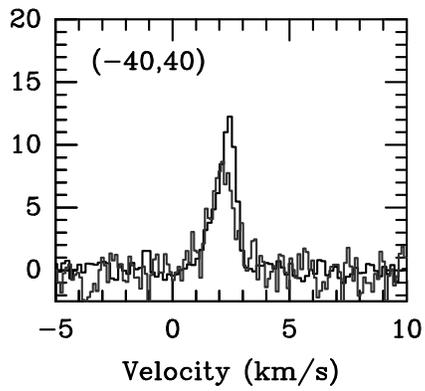
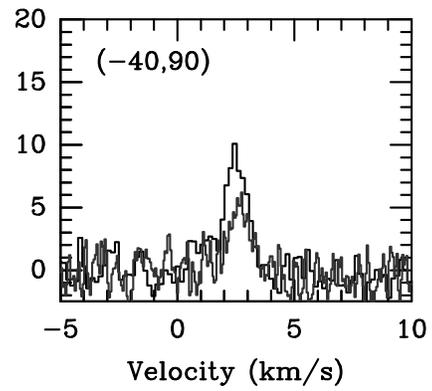
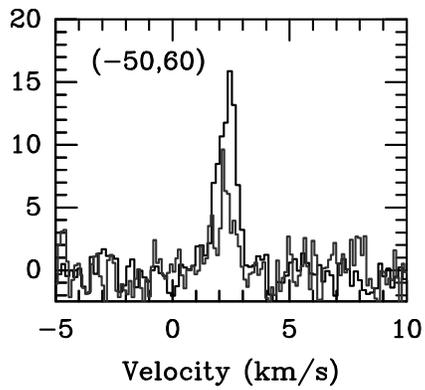
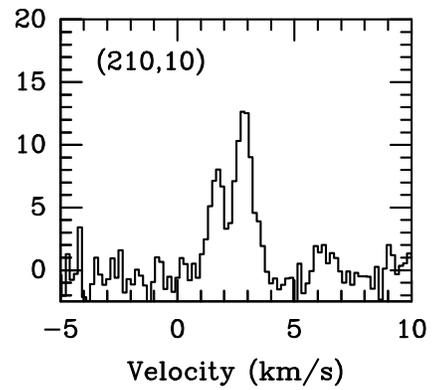
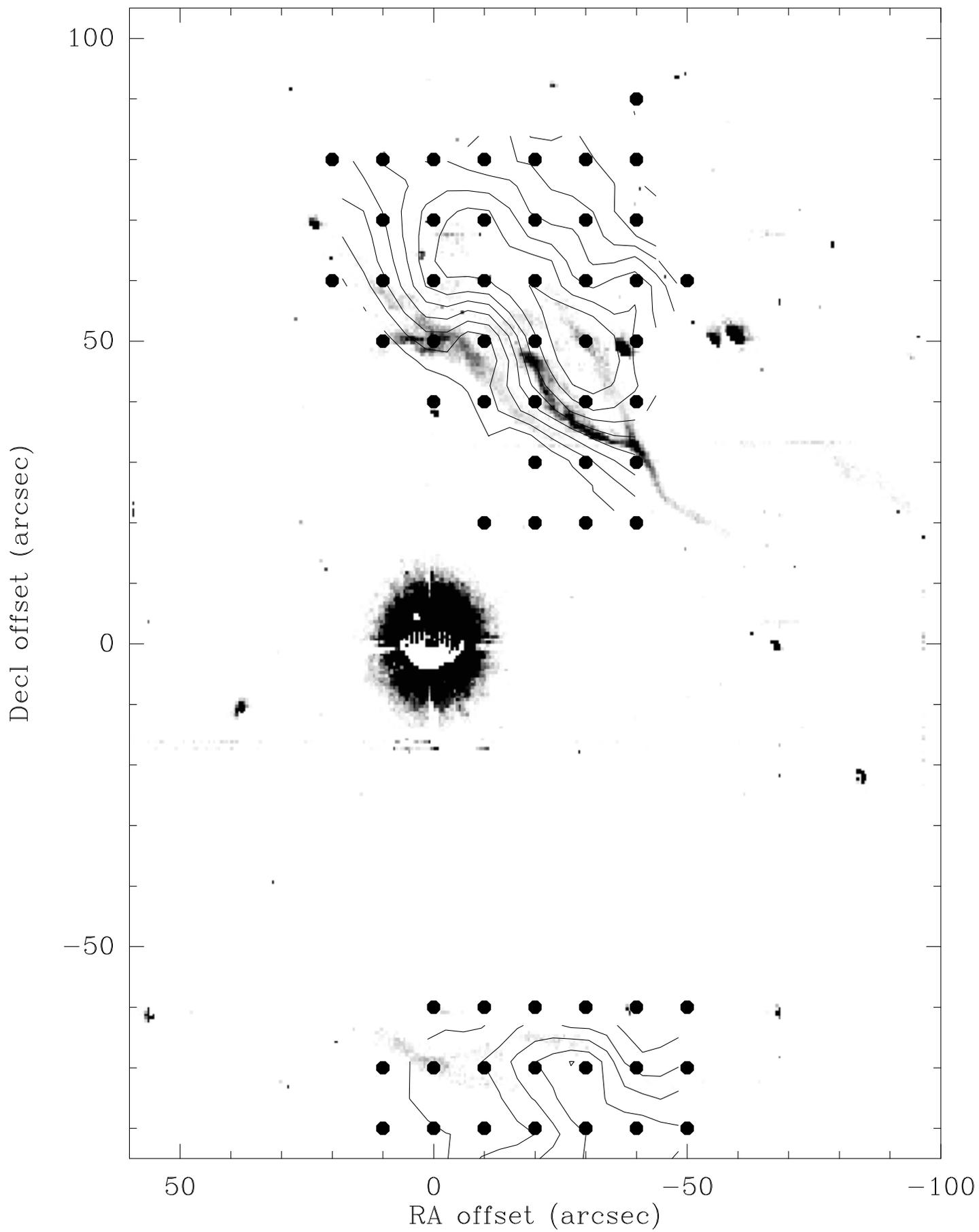
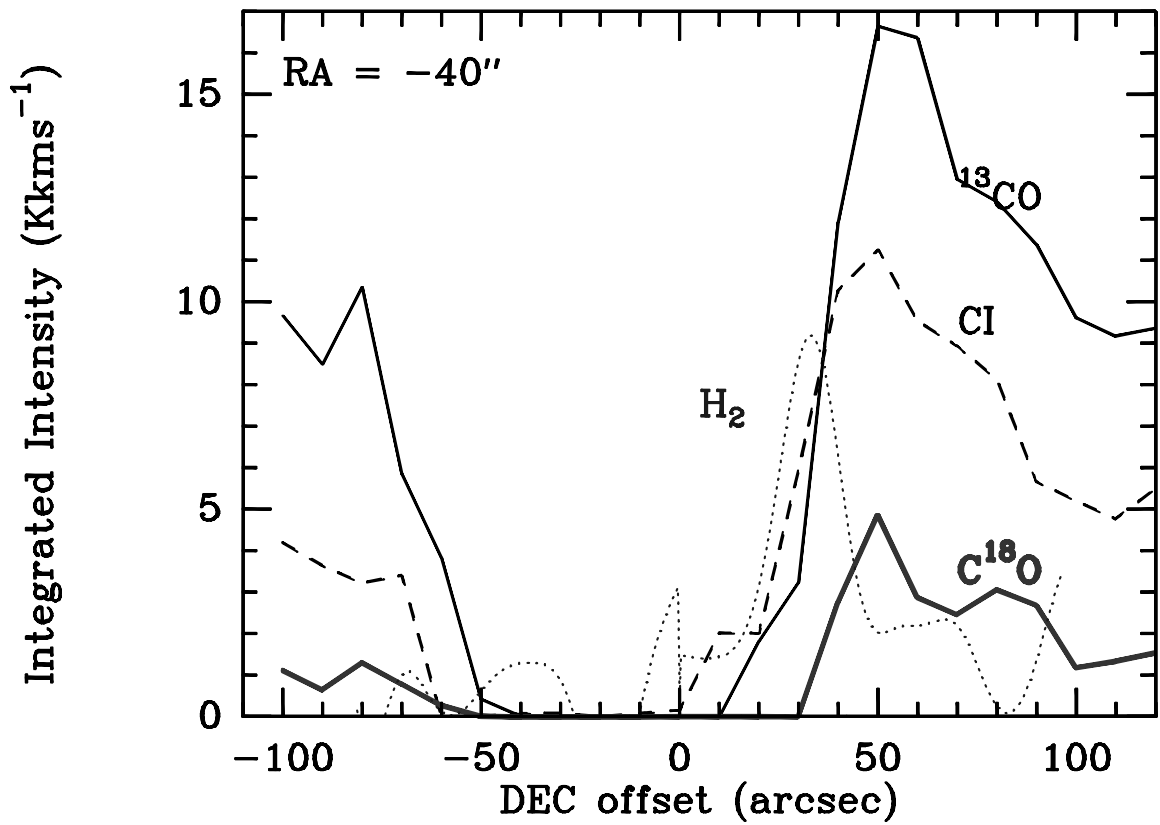
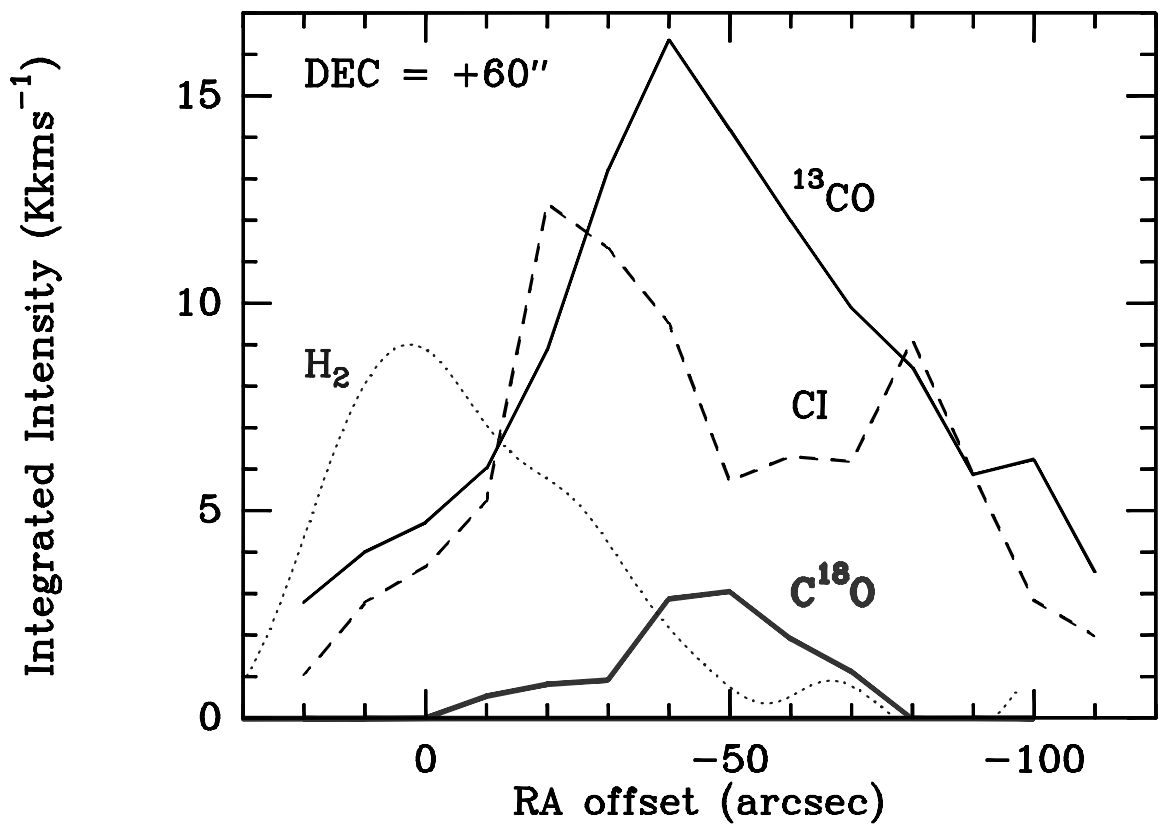
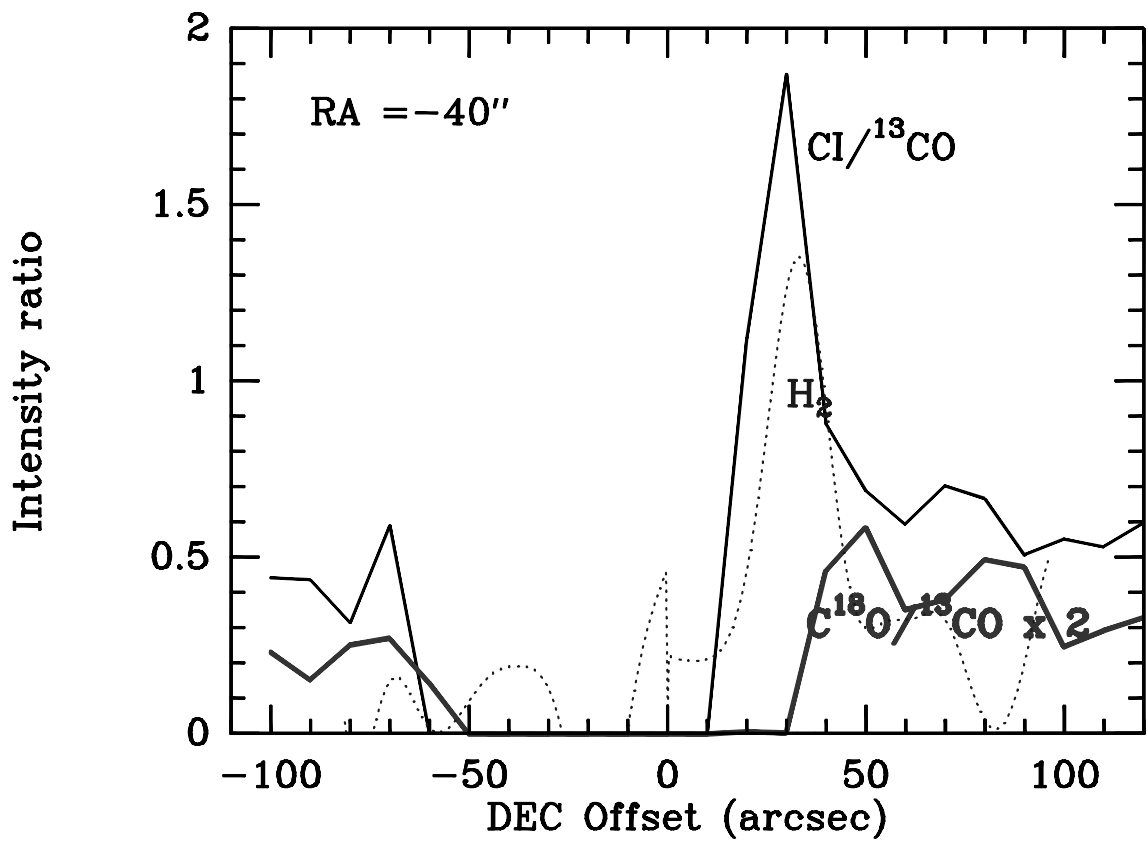
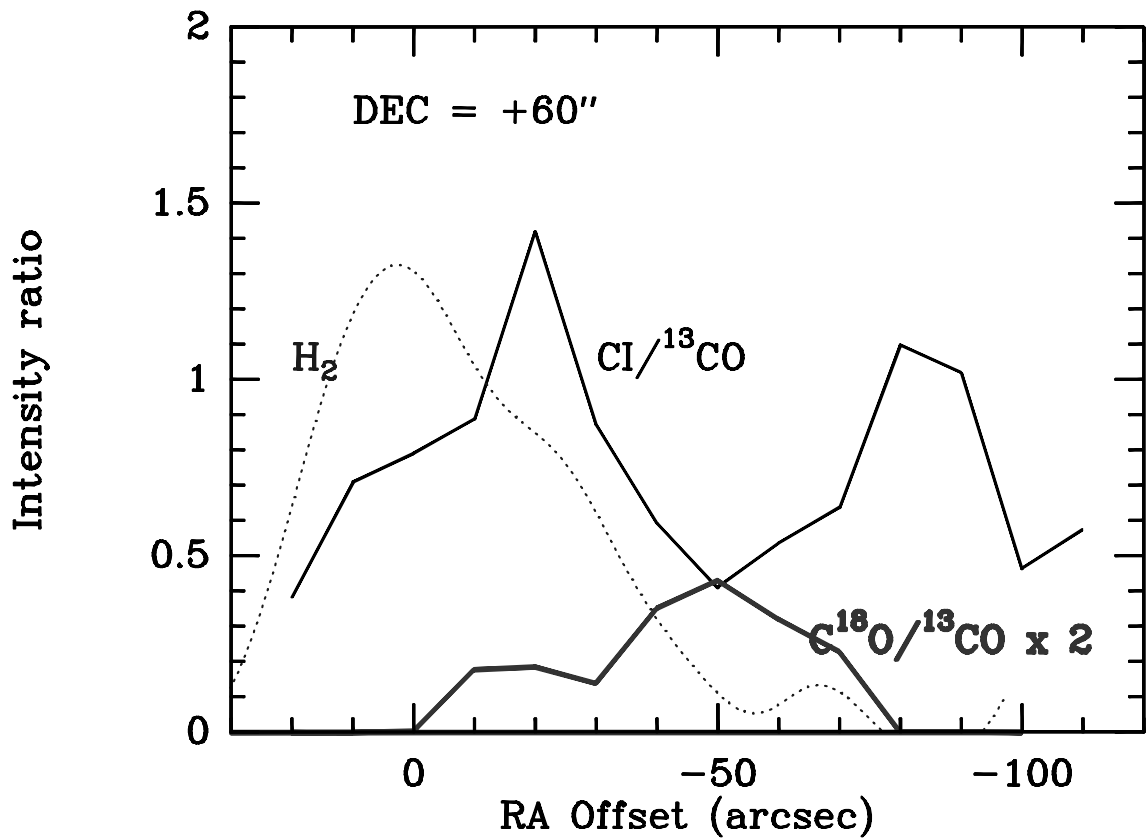
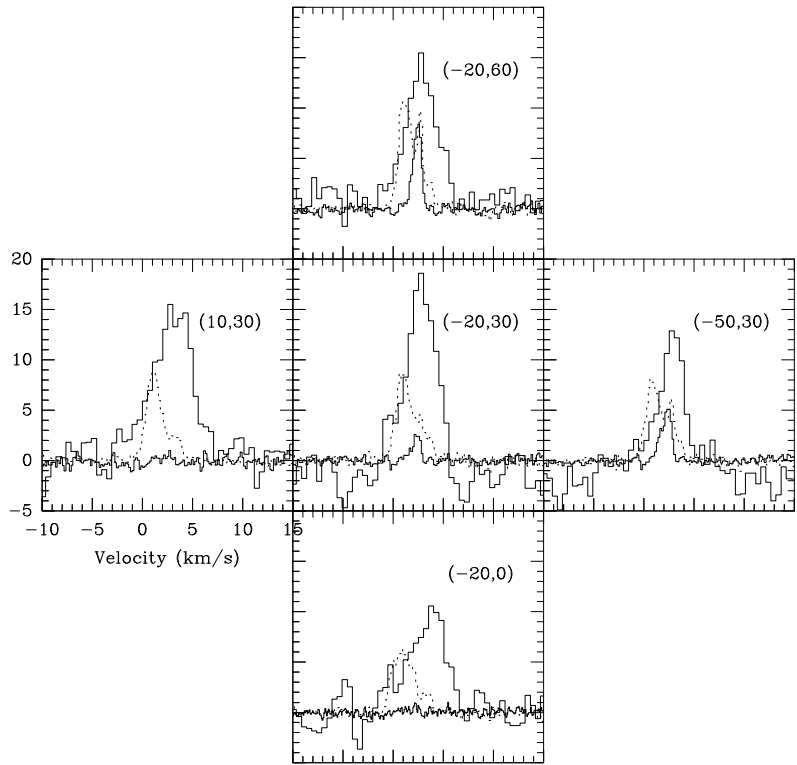
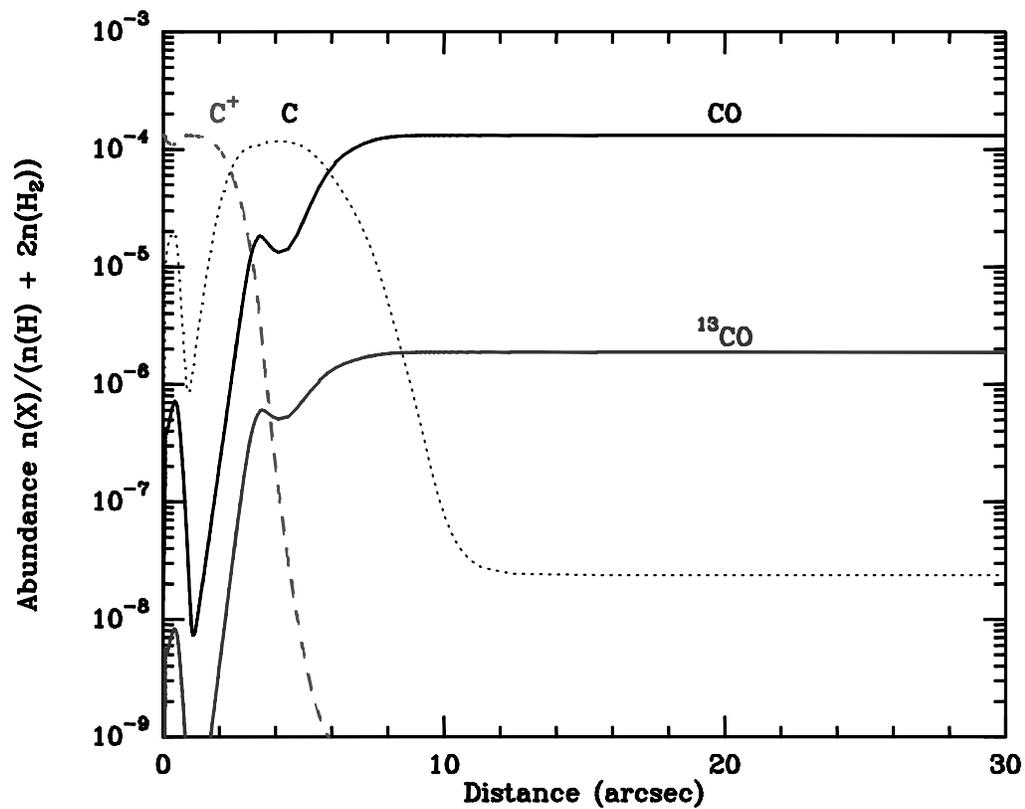
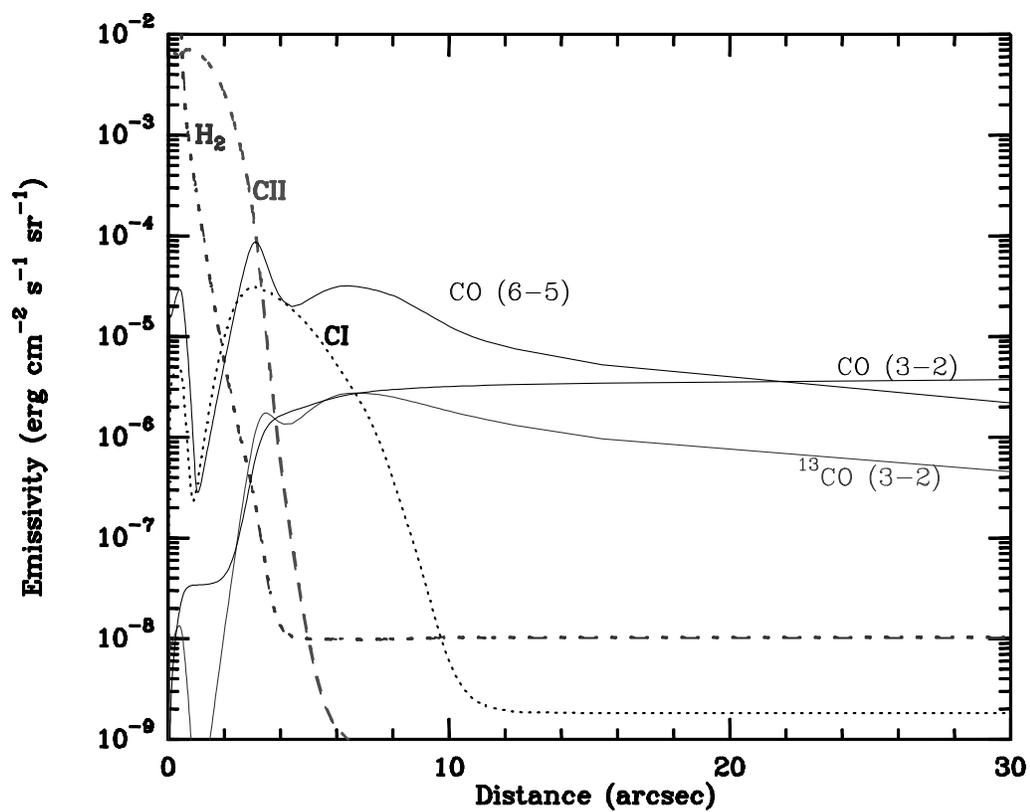
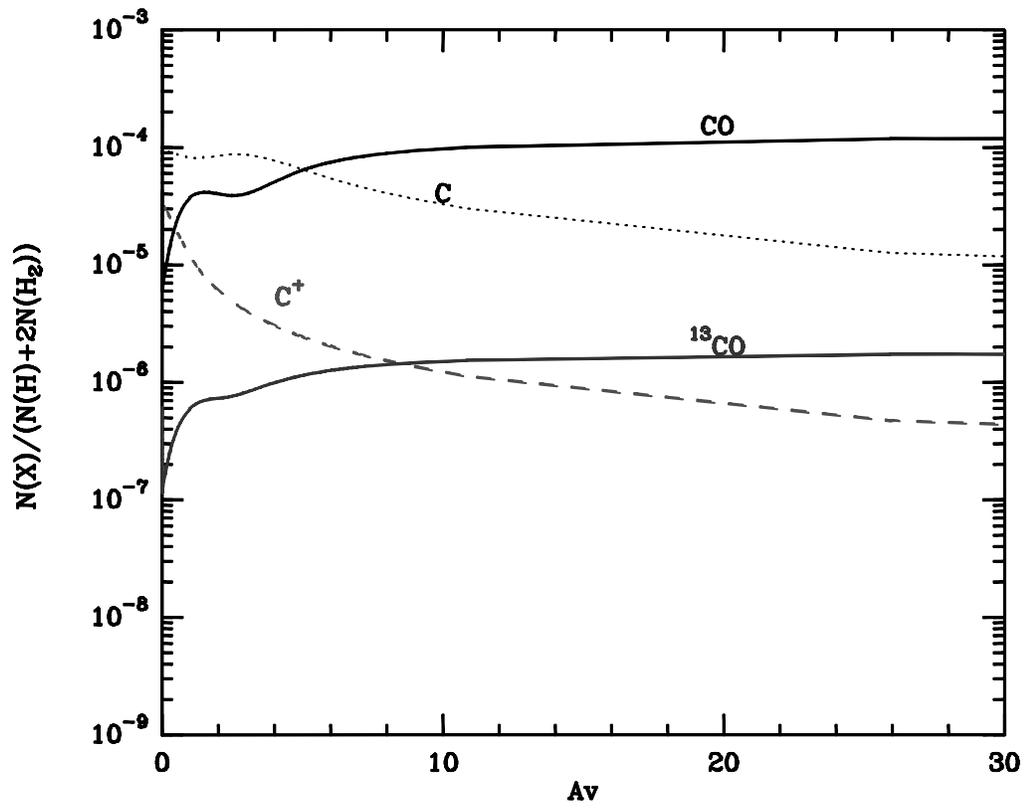
\end{document}